\documentclass[twocolumn,showpacs,preprintnumbers,amsmath,amssymb,prb]{revtex4}

\usepackage{graphicx}
\usepackage{dcolumn}
\usepackage{bm}
\newcommand{\tube} {{\mbox{Na$_{2}$V$_3$O$_7$}}}
\newcommand{\tubex} {{\mbox{Na$_{2-x}$V$_3$O$_7$}}}

\newcommand{\onethree}{{\ensuremath{\frac{1}{3}}}}
\newcommand{\twothree}{{\ensuremath{\frac{2}{3}}}}
\newcommand{\DSCmag} {{\ensuremath{\frac{d^{2}\sigma_{mag}}{d\Omega d\omega}}}}

\newcommand{\DSCSFx} {{\ensuremath{\frac{d^{2}\sigma^{SF}_{x}}{d\Omega d\omega}}}}
\newcommand{\DSCSFy} {{\ensuremath{\frac{d^{2}\sigma^{SF}_{y}}{d\Omega d\omega}}}}
\newcommand{\DSCSFz} {{\ensuremath{\frac{d^{2}\sigma^{SF}_{z}}{d\Omega d\omega}}}}
\newcommand{\DSCNSFx} {{\ensuremath{\frac{d^{2}\sigma^{NSF}_{x}}{d\Omega d\omega}}}}
\newcommand{\DSCNSFy} {{\ensuremath{\frac{d^{2}\sigma^{NSF}_{y}}{d\Omega d\omega}}}}
\newcommand{\DSCNSFz} {{\ensuremath{\frac{d^{2}\sigma^{NSF}_{z}}{d\Omega d\omega}}}}

\begin{document}
\preprint{APS}

\title{New structural and magnetic aspects of the nanotube system ${\tubex}$}
\author{O. Zaharko\cite{now}, J. L. Gavilano, Th. Str\"{a}ssle}
\affiliation{Laboratory for Neutron Scattering, ETH Zurich \& Paul Scherrer Institut, CH-5232 
Villigen, Switzerland}
\author{C. F. Miclea}
\affiliation{Max-Planck-Institute for Chemical Physics of Solids (MPICPfS), D-01187 Dresden, Germany}
\author{A. C. Mota}
\affiliation{Laboratory for Solid State Physics, ETH-H\"{o}nggerberg, CH-8093 Z\"{u}rich, Switzerland}
\affiliation{Max-Planck-Institute for Chemical Physics of Solids (MPICPfS), D-01187 Dresden, Germany}
\author{Y. Filinchuk, D. Chernyshov}
\affiliation{Swiss-Norwegian Beamlines, ESRF, F-38042 Grenoble Cedex 9, France}
\author{P. P. Deen}
\affiliation{Institut Laue-Langevin, 156X, F-38042 Grenoble Cedex 9, France}
\author{B. Rahaman, T. Saha-Dasgupta}
\affiliation{S.N. Bose National centre for Basic Sciences,
JD Block, Sector 3, Salt Lake City, Kolkata 700098, India}
\author{R. Valent\'{\i}}
\affiliation{Institute of Theoretical Physics, University of Frankfurt, D-60438 Frankfurt, Germany}
\author{Y. Matsushita, A. D\"{o}nni, H. Kitazawa}
\affiliation{National Institute for Materials Science (NIMS), Tsukuba, Ibaraki 305-0047, Japan}

\date{\today}

\begin{abstract}
We present new experimental results  of low temperature x-ray synchrotron diffraction, neutron scattering and very low temperature (mK-range) bulk measurements on the nanotube system ${\tube}$.
The crystal structure determined from our data is similar to
the previously proposed model (P. Millet {\it et al.} J. Solid State Chem. $\bf{147}$, 676 (1999)), but also deviates from it in significant details. The structure comprises nanotubes along the $c$-axis formed by stacking units of two V-rings buckled in the $ab$-plane. The space group is P$\overline{3}$ and the composition is nonstoichiometric, ${\tubex}$, x=0.17. 
The thermal evolution of the lattice parameters reveals anisotropic lattice compression on cooling. Neutron scattering experiments monitor a very weak magnetic signal at energies from -20 to 9 meV.
New magnetic susceptibility, specific heat  measurements and decay of remanent magnetization in the 30 mK - 300 mK range reveal that the previously observed transition at  $\approx$76 mK is spin-glass like 
with no long-range order. Presented experimental observations do not support 
 models of isolated clusters, 
but are compatible with a model of odd-legged S=1/2 spin tubes possibly segmented into fragments with different lengths.\\
\end{abstract}

\maketitle

\section{Introduction}
The perspectives to create magnetic and electronic devices on the atomic scale motivates nowadays research on spin clusters, chains, ladders and other geometrical spin arrangements. 
Systems that accommodate such low-dimensional spin arrangements in a periodic fashion, i.e. on a lattice, are of particular interest. This is because their properties arising from individual objects and from the whole assemble can be probed by experimental and theoretical techniques developed for crystalline solids.\\
One of the rare realizations of a crystalline system formed by nanotubes with periodic arrangement is ${\tube}$, a compound first synthesized by Millet {\it et al.}\cite{Millet99}. The structure is formed by VO$_5$ square pyramids connected via edges and apices (Fig.~\ref{fig1}a). There are four inequivalent sodium sites. One Na site is located in the middle of the nanotube and the other three around the nanotubes. Buckled rings formed by nine V atoms (only three sites are inequivalent) have two different  orientations
forming slices A and B (Fig.~\ref{fig1}b). The nanotube is formed by an -A-B-A-B- arrangement along the $c$-axis.\\
Extensive experimental investigations employing specific heat, ac susceptibility and $^{23}$Na nuclear magnetic resonance by Gavilano {\it et al.}\cite{Gavilano03, Gavilano05} revealed puzzling magnetic properties. Those results implied a gradual reduction of the number of V$^{4+}$ S=1/2 magnetic moments, which may be due to the formation of dimers with most of them in a singlet-spin
and a small fraction with a triplet-spin ground state. Presence of a wide range of distributions of singlet-triplet energy gaps of dimerized V moments was claimed.
The total gap distribution consists at least of three groups. Two of them, relatively narrow, are centered at $\approx$ 1 and 10 K and involve only 1/9 of vanadium moments. The third group, that corresponds to the 8/9 part of V moments, is very broad and this leads to a wide range of spin couplings.
Additionally at very low temperatures, $\approx$86 mK, a phase transition was observed in the temperature dependence of the ac-magnetic susceptibility $\chi_{ac}$(T). It was interpreted as a proximity to a quantum critical point at zero field.\\
Several theoretical calculations of the electronic structure and exchange interactions in ${\tube}$ delivered conflicting results.
A spin dimer analysis by Whangbo and Koo\cite{Whangbo00} lead to the conclusion that  ${\tube}$ could be described by six mutually intersecting helical chains. A single helical chain has a spin gap and it is a good approximation for magnetic properties of the whole system. This is not supported by experimental
results. The $\it{ab~initio}$ Density Functional
Theory (DFT) calculations by Saha-Dasgupta {\it et al.}\cite{Dasgupta05}, used the N-th order muffin-tin-orbital (NMTO)
based downfolding method to obtain a 
tight-binding Hamiltonian for V-3d$_{xy}$ orbitals. Their microscopically derived hopping integrals resulted in antiferromagnetic (AF, J$<$0) intra-ring exchange integrals dominating the inter-ring ones. This lead to the description of the system as weakly coupled nine-site rings with J$_1$= -200 K and J$_2$= -140 K intra-ring couplings. 
 Mazurenko {\it et al.}\cite{Mazurenko06} performed
DFT calculations with the local density approximation (LDA)+$U$
approach and suggested that due to the strong hybridization between filled and vacant 3d vanadium orbitals the exchange interaction between V ions is dominated by strong ferromagnetic (F) inter-ring contributions,
while the intra-ring exchange interactions are mainly antiferromagnetic. Their conclusion is that the resulting Heisenberg model is strongly frustrated and there is no simple way to predict the properties of the system. \\
The main goal of our paper is to provide, by characterizing static and dynamic, nuclear and magnetic, microscopic and macroscopic responses,  a solid experimental background necessary to derive a correct model of the magnetic exchange in ${\tube}$.
 
\section{Experimental details}  \label{1}
The ${\tube}$ polycrystalline  samples used in this study have been prepared by a solid-state reaction method using Na$_4$V$_2$O$_7$, V$_2$O$_3$ and V$_2$O$_5$ as described by Niitaka {\it et al.}\cite{Niitaka02}.
The precursor Na$_4$V$_2$O$_7$ was synthesised from a mixture of NaCO$_3$ and V$_2$O$_5$ with a stoichiometric ratio, on a
gold sheet at 873 K under air, and kept at 473 K before use. V$_2$O$_3$ was reduced from V$_2$O$_5$ by 7\% H$_2$ + 93\%
Ar mixture gas. Under dry- and pure-Ar atmosphere, the stoichiometric amount of Na$_4$V$_2$O$_7$+V$_2$O$_3$+V$_2$O$_5$ was
well mixed, and pressed into pellets. The pellets were wrapped in gold foils, and loaded into a fused silica
tube. The tube was subsequently sealed under vacuum and heated in a furnace at 978 K for a week. The product was checked as pure ${\tube}$ by x-ray powder diffraction\cite{note0}.\\
Analysis of the crystal structure and its evolution with temperature was studied by x-ray single crystal and powder diffraction.
A black fiber-like single crystal (10x10x700 $\mu$m) was extracted from the polycrystalline sample.
X-ray diffraction studies have been performed at the BM01A Swiss-Norwegian Beamline (SNBL) of ESRF, France, at a wavelength of 0.722870 \AA, and at the Material Science Beamline (MSB) of SLS, Paul Scherrer Institute, Switzerland, with $\lambda$=0.620967 \AA.
The crystal structure has been determined from a single crystal at SNBL at three different temperatures, 200 K, 80 K and 16 K, using an image-plate area detector MAR345 and N$_2$ or He flow cryostats to reach low temperatures. 180 deg $\omega$ oscillation images with an
increment of 1 deg were collected with 1 min of exposure and
the crystal-to-detector distance of 180 mm (see Table~\ref{tab1}). 
The quality of the data at 16 K is lower than at 80 K and 200 K due to a lower resolution obtained with the complex low-temperature experimental setup. 
The data were corrected for the Lorentz-polarisation factor and for absorption. For the structure solution and refinements the programs SHELXS and SHELXL\cite{SHELX} were used.\\
Powder x-ray patterns were collected in a 0.5 mm quartz capillary at temperatures between 80 K - 470 K  with the same setup used for the single crystal measurements. The experiment was performed on heating, with a heating rate of 120 K/h. One exposure lasted 30 s and the capillary was rotated 30 deg/exposure for better powder average. For patterns recorded in the 
temperature range of 14 K - 130 K a Janis He cryostat at MSB and a microstrip detector were used with 30 s exposure. Structure refinements were performed using the Rietveld technique implemented in the Fullprof program\cite{fullprof}.\\
Magnetic excitations and lattice vibrations were investigated on a 10 g polycrystalline sample. Inelastic neutron scattering measurements were performed on the neutron time-of-flight (TOF) spectrometer FOCUS at SINQ, PSI, Switzerland, at 1.5 K and 120 K. Three setups were exploited ($\lambda_i$= 5.75, 4.85 and 1.7 \AA) to access energy-transfer ranges below 0.9, 2 and 20 meV, respectively, with optimal resolution. Contributions from an empty Al sample holder were measured and subtracted.\\
A powder neutron diffraction experiment  with full three-directional (XYZ) neutron polarisation analysis was performed on the D7 instrument~\cite{Stewart2008} at the Institute Laue-Langevin, France. 
The incident neutron wavelength 3.1 \AA~was selected using a pyrolytic graphite monochromator. Polarisation of the incident and analysis of the scattered beam were performed using polarising mirrors. Helmholtz coils enabled rotation of the incident neutron spins along three mutually perpendicular directions.
The efficiency of the 132 detectors  of D7 was determined using a standard vanadium sample, while the polarisation efficiency was determined using a quartz sample with entirely non-spin flip scattering. The measurements were performed at 1.9, 120 K (each 60 h) and 300 K (2 h). 
Our experimental setup allowed us to measure the response of the sample within the wave vector
range 0.5 $<$ Q $<$ 4 \AA$^{-1}$ and to integrate the signal
over the energy interval -$\infty$, 9 meV. The scattering from the sample was brought to the absolute scale using vanadium normalization. Absorption of the vanadium and the sample was taken into account.\\
A powder non-polarized neutron diffraction experiment in the mK range was performed on DMC at SINQ, PSI, Switzerland, using a wavelength of $\lambda$=2.457 \AA. The sample was mounted into the dilution insert of an ILL orange cryostat and two diffraction patterns, at 40 mK and 1 K, were collected.\\
Measurements of the susceptibility on a polycrystalline sample were done at MPICPfS, Dresden in the 30 mK - 3.5 K range in an impedance bridge using a SQUID as a null detector. In this arrangement, the amplitude of the ac field can be varied in fixed steps from 0.07 to 33 mOe and the frequency can be chosen from 4 different values between 16 and 160 Hz. The sample was placed inside the mixing chamber of a dilution refrigerator. The residual field in the cryostat was less than 20 mOe. Specific heat measurements were performed using a quasi-adiabatic method in the
temperature range 60 mK - 4 K at four values of applied field 0, 0.25, 0.5 and 2 T.\\
The decay of the remanent magnetization was measured at MPICPfS, Dresden. The sample was cooled in a field of 100 Oe from T= 200 mK to the desired temperature T$<$T$_f$. Note that 200 mK is  well above the critical temperature T$_f$= 86 mK where a cusp in $\chi_{ac}$ occurs. After a waiting time of about 10 hours, the field was removed and the relaxation of the remanent magnetization was recorded with a flux counter for about 20000 - 70000 seconds. After these times, the sample was heated to 200 mK and the expelled flux was recorded in order to obtain the total value of the remanent magnetization $M_{rem}= M_{0}-M_{\infty}$.\\

\section{Results}  \label{2}

\subsection{Temperature evolution of the crystal structure} \label {21}
\begin{figure}[tbh]
\caption {a) [001] projection of the $\tube$ structure.
Arrangement of V atoms forming buckled rings (slices A, B) within the nanotube b) in the model of Millet {\it et al.}\cite{Millet99} and c) in our model. The three sites of vanadium are denoted by V1 - red, V2 - pink, V3 - blue. Oxygen and sodium atoms are omitted for clarity. d) Four different families of exchange couplings $J_1$ (grey), $J_2$ (white), $J_3$ (black), $J_4$ (black dashed).}
\includegraphics[width=86mm,keepaspectratio=true]{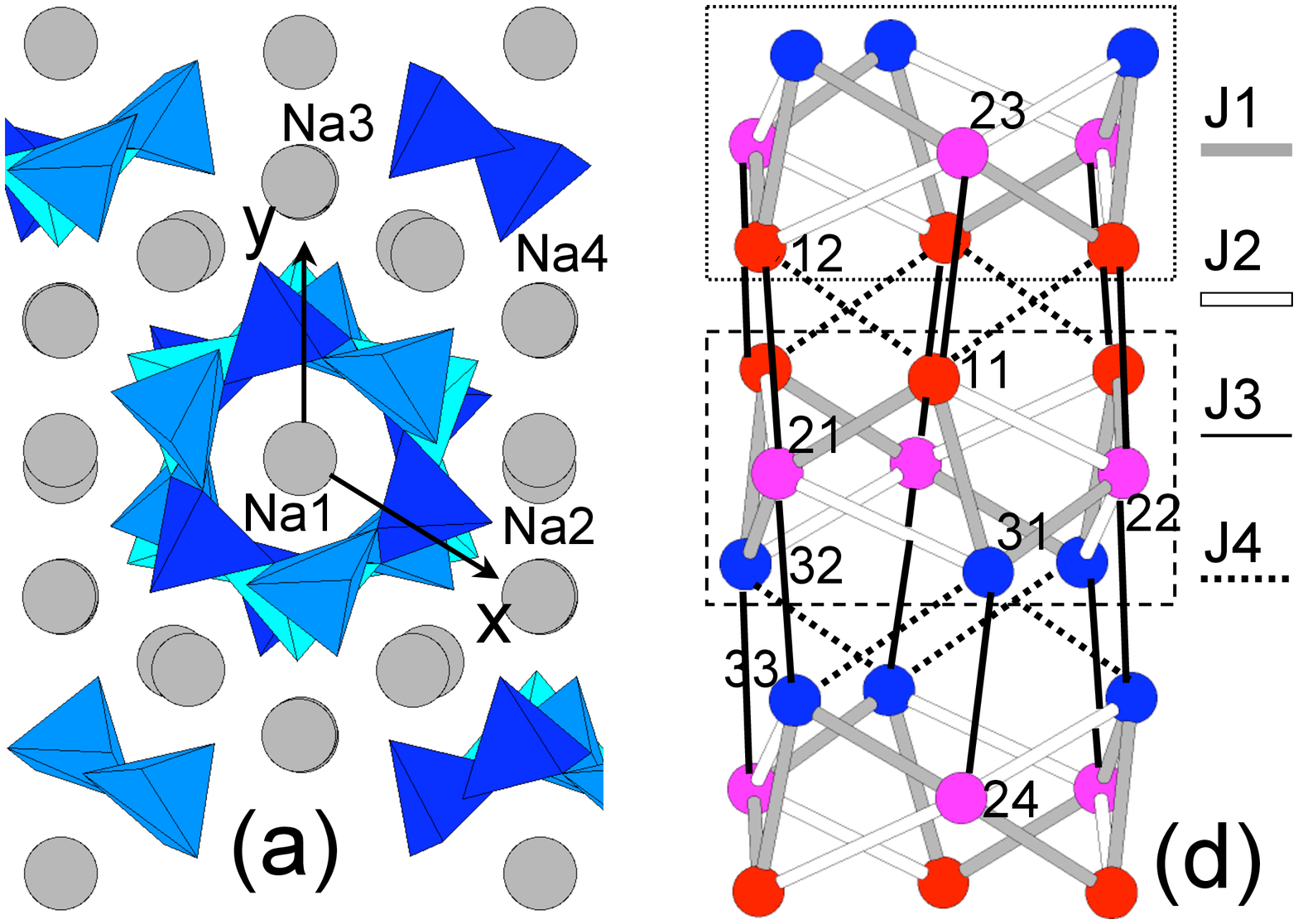}
\includegraphics[width=66mm,keepaspectratio=true]{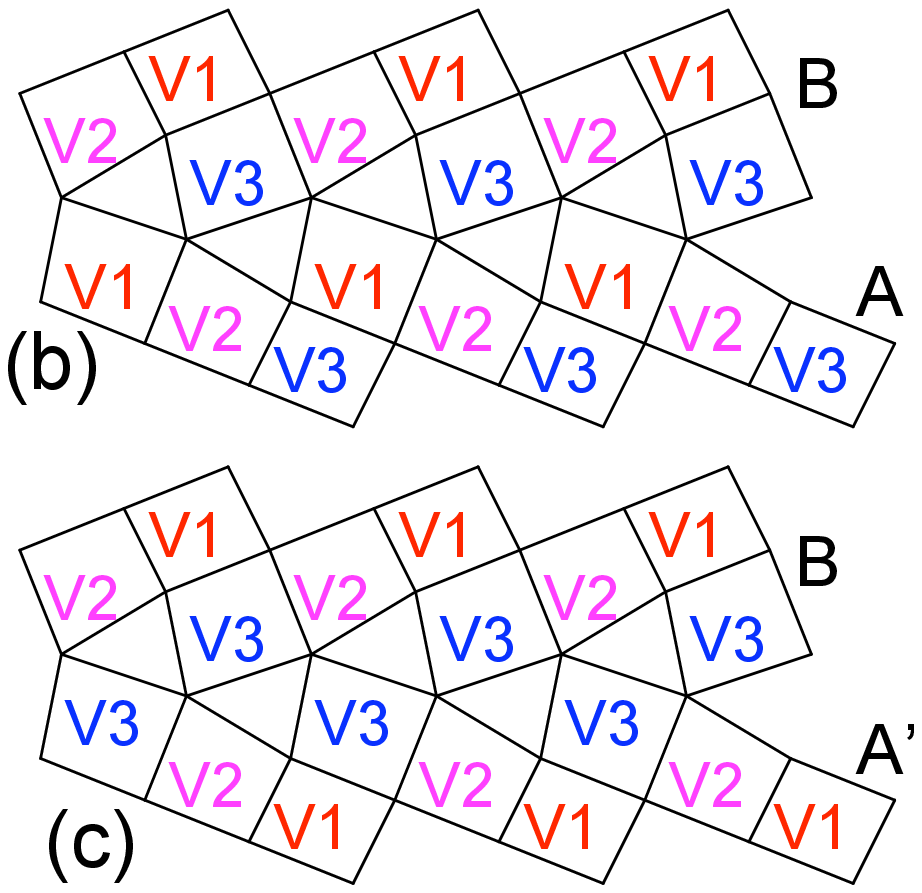}
\label{fig1}
\end{figure}

\begin{figure}[tbh]
\caption {Thermal expansion of the lattice parameters $a$ (red) and $c$ (blue) from x-ray powder diffraction. Inset: The $a/c$ ratio calculated from the SNBL (green), SLS (violete) powder patterns and SNBL (red) single crystal data.}
\includegraphics[width=86mm,keepaspectratio=true]{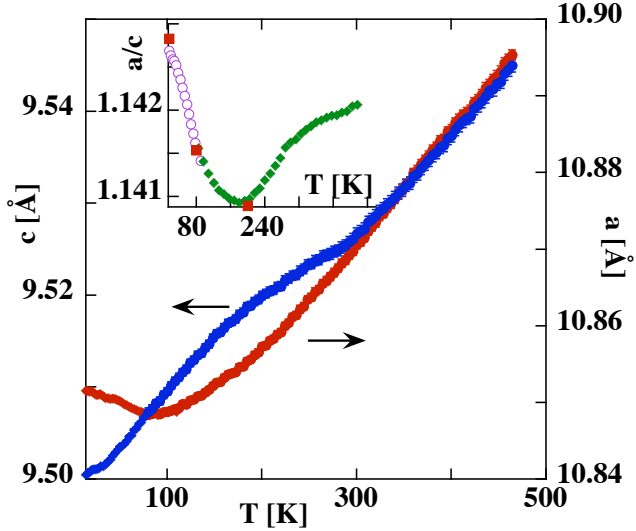}
\label{fig2}
\end{figure}

\begin{figure}[tbh]
\caption {Thermal evolution of the V-V distances calculated from single crystal x-ray diffraction at 200, 80 and 16 K. Top: 1$^{st}$ intra-ring distances, bottom:  2$^{nd}$ intra-ring (green), 1$^{st}$ (red) and 2$^{nd}$ (blue) inter-ring distances.}
\includegraphics[width=86mm,keepaspectratio=true]{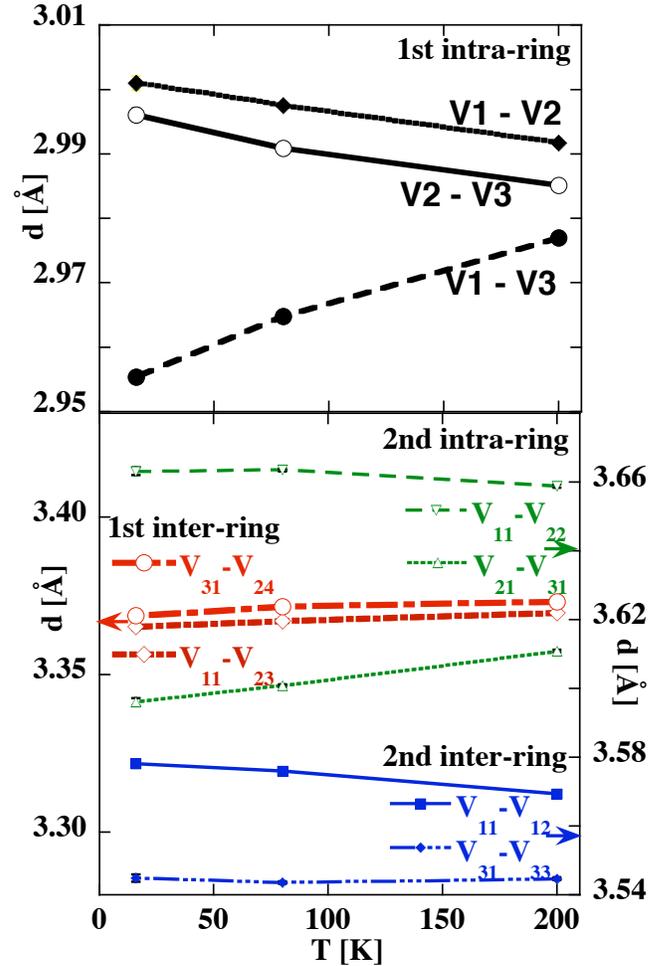}
\label{fig3}
\end{figure}

In an attempt to understand the magnetic exchange in ${\tube}$ we performed detailed structural investigations at low temperatures. The analysis of the 200~K x-ray single crystal diffraction data set suggested an apparent Laue symmetry $\overline{3}m$ with mirror planes situated along the [100], [010] and [110] directions. 
No systematic absences of the space group $P31c$ reported by Millet{\it et al.}\cite{Millet99} were observed.  
The direct methods were tentatively used with the following probable space groups $P\overline{3}1m$,
$P31m$ and $P312$, but no acceptable solution was found. Therefore, the presence of $\{$010$\}$ twinning of the crystal with $P3$ or $P\overline{3}$ symmetry has been considered. The centrosymmetric model presented in Table~\ref{tab2} with 50\% of two twin domains explained well the apparent $\overline{3}1m$ symmetry. The R-values corresponding to the final refinements are presented in Table~\ref{tab1}. The anisotropic displacement parameters and occupancies of all atoms could be refined  at 80 K and 200 K (see Table~\ref{tab2b}). For the 16 K data set, due to its lower resolution, isotropic displacement parameters were used. \\
Single crystal data collected at three temperatures indicate no structural phase transition.
The structure resulting from our analysis is very similar to the model proposed by Millet {\it et al.}\cite{Millet99} However, there are important differences. 
The arrangement of nine V atoms into the buckled ring (slice) B is the same as in the model of  Millet (Figs.~\ref{fig1}b and c). But due to the absence of the glide plane in the $P\overline{3}$ space group, the slice A' is oriented differently compared to A. We identify the A'-B two-ring stacking unit as a structural building block of ${\tube}$.\\
A detailed structural investigation resulted in precise values of the V-V distances and V-O-V angles presented in Tables~\ref{tab3},\ref{tab4}. Here a second index in the notation of V atoms is introduced to simplify the comparison of numerical and graphical presentations (see also Fig.~\ref{fig1}d). Based on these values it is possible to distinguish four different families of effective exchange paths. They are: 1$^{st}$ and 2$^{nd}$ intra-ring and 1$^{st}$ and 2$^{nd}$ inter-ring ones. In a simplified model for the magnetic exchange in $\tube$ each family may be represented by a single exchange coupling ($J_1 -J_4$, see Fig.~\ref{fig1}d). The 1$^{st}$ intra-ring exchange path is characterized by the average shortest V-V distance  of $\approx$2.985~ \AA~and the V-O-V angles of $\approx$99 deg. Thus it is the most important coupling. The 2$^{nd}$ intra-ring path has rather large V-V distances ($\approx$3.63 \AA) and V-O-V angles of $\approx$140 deg. The 2$^{nd}$ inter-ring path has smaller V-V distances ($\approx$3.55 \AA) and almost the same V-O-V angles of $\approx$132 deg. Therefore, based solely on the atomic arrangement, the 2$^{nd}$ intra-ring and 2$^{nd}$ inter-ring paths might be equally important. The 1$^{st}$ inter-ring path will certainly result in a 'separate' coupling as the V-V distance is $\approx$3.37 \AA~and the V-O-V angles are $\approx$119 deg. 
DFT calculations\cite{Dasgupta05} performed on the crystal structure proposed by Millet {\it et al.}\cite{Millet99} identify the 1$^{st}$ and 2$^{nd}$ intra-ring couplings as the most important ones.  And, according to these calculations, the 2$^{nd}$ intra-ring and 2$^{nd}$ inter-ring couplings are significantly different  due to the orientation of the V orbitals.
As small differences in the atomic arrangement between our model and the model of Millet, can significantly change the strength and even the sign of the magnetic exchange\cite{Mazurenko06}, we revised the electronic structure of $\tube$ performing new DFT calculations (see Section~\ref{3}).\\

In the new model of crystal structure, within the vanadium-oxide nanotubes there are two inequivalent Na sites, 1(a) and 1(b). They are only partially occupied. This leads to a nonstoichiometric composition Na$_{2-x}$V$_3$O$_7$ with x=0.17. The requirement of neutrality of the
crystal inquires that the missing positive charge is distributed on the neighboring atoms; either some amount of V$^{5+}$ (up to 5.7\%) or intermediate valence states should be present. These minority states, if localized on V, could serve as defects introducing randomness into the V-V exchange couplings. \\
The displacement parameters U of all atoms, including sodium, are small, almost isotropic (see Table~\ref{tab2b}) and
decrease with temperature lowering, as in conventional materials. Therefore, the assumptions used to identify the low frequency phonon mode at 88 cm$^{-1}$ as the Na rattling mode\cite{Choi02} does not hold. Our Na$^{+}$ mean square displacement amplitude $<u^2>$ is approximately six times smaller than assumed by Choi et al.\cite{Choi02} and therefore the characteristic frequency would be  $\approx$ 174 cm$^{-1}$, which is incompatible with the observed hypothesized rattling mode at 88 cm$^{-1}$.\\
One more important outcome of our x-ray diffraction study is the temperature evolution of the lattice parameters $a$ and $c$. Powder patterns reveal that the lattice compresses gradually on cooling (Fig.~\ref{fig2}). This compression is very anisotropic, as visualized in the $a/c$ ratio in the inset of Fig.~\ref{fig2}. Whilst above 300 K the expansion coefficients 
$\Delta a/a$ and $\Delta c/c$ lie in the range of 10$^{-6}$/K, which is typical for oxides; several anomalous points in $a$(T) and $c$(T) are observed at lower temperatures. Near 300 K the decrease of $c$ slows down, close to 100 K with decreasing temperature, $a$(T) starts to increase. Unfortunately we cannot supply microscopic details of all these changes, but we reckon that the changes near 100 K correlate with the anomaly observed in the $^{23}$Na-NMR spin-lattice relaxation rate\cite{Gavilano05} and the 88 cm$^{-1}$ mode detected in optical measurements\cite{Choi02}.\\
The variation of the V-V distances and V-O-V angles obtained from single crystal refinements (Tables~\ref{tab3},\ref{tab4} and Fig.~\ref{fig3}) allows us to conclude that the dominant tendency is a reduction of the V-V distances and V-O-V angles having the largest projection along the tubes and an increase of those having a substantial xy-component.\\

\subsection{Neutron scattering: polarized diffraction and time-of-flight inelastic scattering}  \label{22}

We performed neutron scattering experiments to obtain information on the static and dynamic magnetic correlation functions in ${\tubex}$. 
This system has significant incoherent scattering $\sigma_{incoh}$ due to the presence of  vanadium and sodium. $\sigma_{incoh}$ is comparable to the magnetic scattering of V ($\sigma_{incoh}$=1.47 barn/str. f.u.,
$\sigma_{mag}$=0.437 barn/str. f.u.), additionally absorption is significant. Therefore, a polarised neutron diffraction experiment was carried out  on D7 to separate magnetic scattering from
nuclear coherent, nuclear spin incoherent and background contributions.
The paramagnetic scattering was obtained by averaging the contributions extracted from the spin-flip (SF) and non-spin-flip (NSF) partial cross sections\cite{Scharpf}:
\begin{eqnarray}
{\DSCmag} &=& 2 [{\DSCSFx} + {\DSCSFy} - 2{\DSCSFz}]
\nonumber\\
{\DSCmag} &=& 2 [-{\DSCNSFx} - {\DSCNSFy} + 2{\DSCNSFz}]
\label{crossec}
\end{eqnarray}
where z is perpendicular to the scattering plane.
\begin{figure}[tbh]
\caption {Top: total, nuclear spin incoherent  and magnetic scattering at 120 K measured on D7 with $\lambda_i$=3.1 \AA. Inset: The (Q, $\omega$) window accessible during the experiment is shown by the dashed area. Bottom: Absolute magnetic scattering at 1.8, 120 and 300 K.}
\includegraphics[width=86mm,keepaspectratio=true]{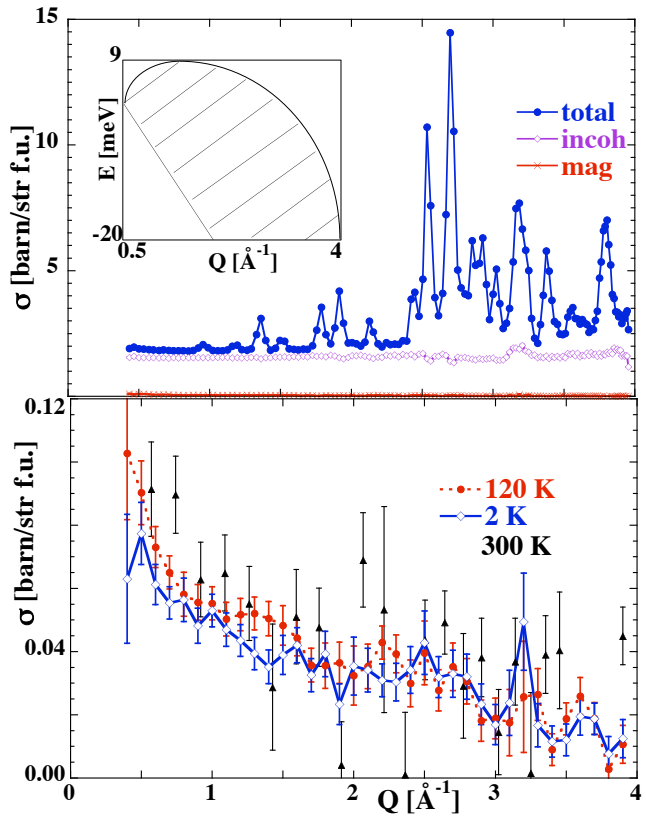}
\label{d7}
\end{figure}
\begin{figure}[tbh]
\caption {Comparison of the magnetic scattering at 2 K measured on D7 (red) and the calculated paramagnetic signal of V$^{4+}$ (black solid), both in absolute scale. Calculated low-lying excitations of an isolated nine-member ring (violet dashed) and of an isolated dimer (blue dotted) are given in arbitrary units.}
\includegraphics[width=86mm,keepaspectratio=true]{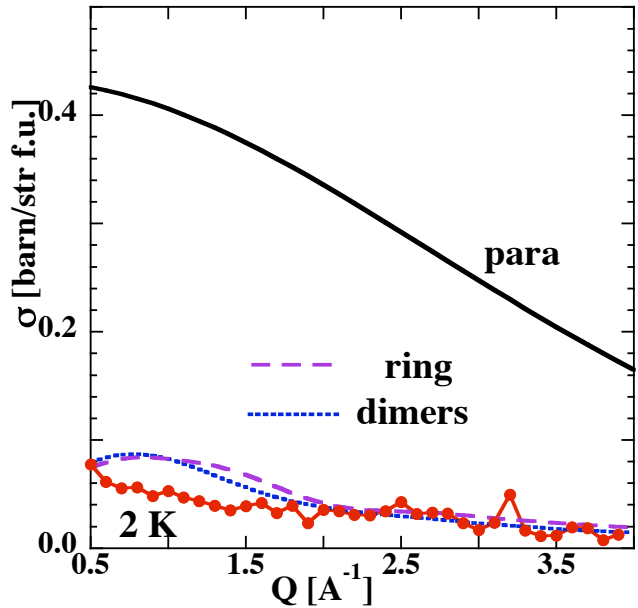}
\label{INSclusters}
\end{figure}
\begin{figure}[tbh]
\caption {INS spectra of  ${\tubex}$ at 1.5 K (blue) and 120 K (red) measured on FOCUS with $\lambda_i$=1.7 \AA. The intensity has been integrated over the momentum transfer range 2.1 $<$ Q $<$ 5.7 \AA$^{-1}$. Inset: Q-dependence at 120 K for 2.6 $<$ Q $<$ 3.4 \AA$^{-1}$ (green) and 4.1 $<$ Q $<$ 4.9 \AA$^{-1}$ (magenta).}
\includegraphics[width=86mm,keepaspectratio=true]{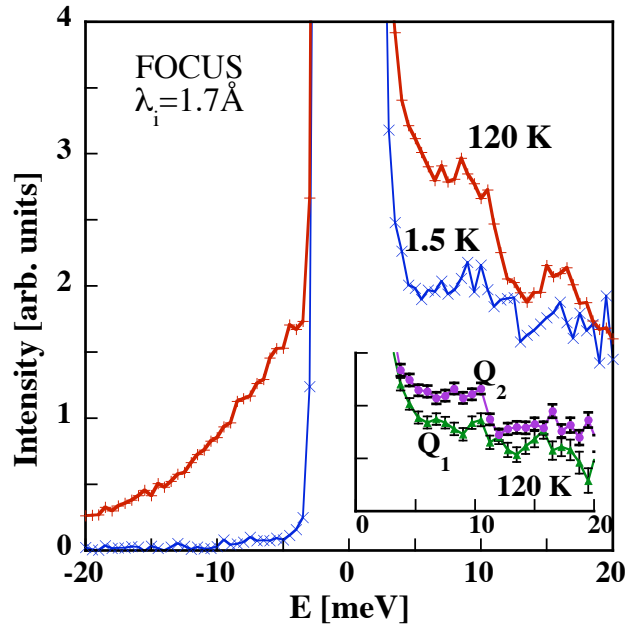}
\label{focus}
\end{figure}
Fig.~\ref{d7} (top) presents total, nuclear spin incoherent and magnetic signals from the sample. Apparently the magnetic signal is tiny. 
The amount of magnetic scattering summed over the measured Q-range is only 13 \% of the expected paramagnetic scattering of ${\tubex}$ formula unit (f.u.) in this Q-interval calculated as  
\begin{equation}
\sigma_{mag} = \Sigma_{Q} {\twothree}NS(S+1)(r_o \gamma)^2 f^2(Q),
\label{mag}
\end{equation}
where ${\twothree}$ results from the powder average, $N$=3 is the number of spin $S$=1/2 V$^{4+}$ ions per f.u.,
$(r_o \gamma)$ = -0.54$\cdot$10$^{-12}$ cm and $f(Q)$ is the magnetic form factor of V$^{4+}$.
The weakness of the signal caused long counting times to obtain reasonable statistics in the diffraction mode, i.e. integrating over all energy window, and it was impossible to measure the (Q, $\omega$) distribution of the magnetic scattering in TOF mode of D7.\\
Fig.~\ref{d7} (bottom) presents magnetic contributions at 1.8, 120 and 300 K. The signals are almost equal and featureless within statistical errors. This leads to important conclusions. Firstly, the majority of magnetic intensity remains out of the integrated energy and momentum window of our experiment shown as an inset in Fig.~\ref{d7} (top). At 1.8 K dominantly elastic and neutron energy loss events are measured, while at 120 K additionally the neutron energy gain side up to $\approx 2kT$=20 meV is accessible. In both measurements the magnetic scattering is very weak and it only slightly increases from 1.8 to 120 K in the low Q range, which implies presence of high-energy states becoming populated at higher temperature. The
total paramagnetic scattering is not recovered even at 300 K (Figs.~\ref{d7}, \ref{INSclusters}), therefore, most of the magnetic intensity is located at even higher energy transfers.\\
Secondly, observation of weak magnetic intensity with no oscillation in the Q-dependence disputes the presence of isolated clusters. Any isolated magnetic cluster would have a characteristic oscillation in the Q-dependence, which arises due to the geometric $\frac{sin(Q \Delta R)}{Q \Delta R}$ term in the neutron cross section\cite{Furrer79}:
\begin{eqnarray}
{\DSCmag} \propto f^2(Q)\Sigma^n_{j<j'} (|<\bf{S}||\bf{T_j}||\bf{S'}>|^2)+
\nonumber\\
2\frac{sin(Q \Delta R)}{Q\Delta R}<\bf{S}||\bf{T_j}||\bf{S'}><\bf{S'}||\bf{T_{j'}}||\bf{S}>
\label{GeoTerm}
\end{eqnarray}
where $\Delta R= |\bf{R_j}-\bf{R_{j'}}|$ is the distance between the V$^{4+}$ ions $i$ and $j$ of the cluster, $\bf{T_j}$ - are irreducible tensor operators.
Fig.~\ref{INSclusters} presents the expected Q-dependences for an isolated nine-member ring and an isolated dimer calculated using
programs of Weihe\cite{Woihe}. For the nine-member ring the Hamiltonian is $H = J_1\Sigma^9_{i=1}({\bf{S}}_i\cdot {\bf{S}}_{i+1}) + J_2\Sigma_{i,j}({\bf{S}}_i\cdot {\bf{S}}_{j})$ (for $i$=2 $j$=4, 9, for $i$=5 $j$=3, 7, for $i$=8 $j$=1, 6) and the two intra-ring antiferromagnetic exchange constants are $J_1$=-160 K and $J_2$=-90 K, the best estimate from our present susceptibility data (see Section~\ref{4}).

The geometric term is summed over the excitations between the 16 lowest cluster levels which could contribute to the integrated energy window of the D7 experiment and $\Delta R$'s correspond to the V-V intra-ring distances determined in this work. For the dimer the Hamiltonian is $H = J(\bf{S}_1\cdot\bf{S}_2)$ with the antiferromagnetic exchange constant $J$  chosen such that the spin singlet-triplet excitation lies
in the energy window of the experiment, $\Delta R$ is the average 1$^{st}$ intra-ring distance.
None of the calculated Q-dependences is consistent with the measured one. \\
Yet, our observations are compatible with two following pictures: i) $\tubex$ is composed of complex objects, i.e. spin clusters with different sizes and exchange couplings, and the Q-dependence of each individual cluster is blurt out; ii) nine-member ring spin clusters are coupled into tubes, this leads to dispersion of excitations and to smearing of oscillations in the measured energy integrated S(Q, $\omega$). 
An intermediate picture, nanotubes segmented into finite fragments with different length, also explains the observations.\\
Inelastic neutron scattering collected at two temperatures 1.5 and 120 K  (Fig.~\ref{focus}) in the TOF mode on FOCUS is dominated by incoherent scattering. No distinct magnetic excitations have been observed up to 20 meV in agreement with the D7 results. The weak  mode observed
near 0 meV is most probably due to phonons since its intensity increases with Q and temperature (Fig.~\ref{focus} inset). It is found close to the 88 cm$^{-1}$ mode observed by optical measurements\cite{Choi02}. Our observations, however, does not allow us to specify which atoms (V or Na) are involved in this mode.

\subsection{Magnetic properties in the mK temperature range}  \label{23}
\subsubsection{ac-susceptibility and specific heat}

We performed new experiments in the mK temperature range to characterize the previously found~\cite{Gavilano05} very low-temperature state of ${\tubex}$.
\begin{figure}[tbh]
\caption {Temperature and frequency dependence of real $\chi'$ and imaginary $\chi''$ components of ac-susceptibility in mK temperature range.}
\includegraphics[width=86mm,keepaspectratio=true]{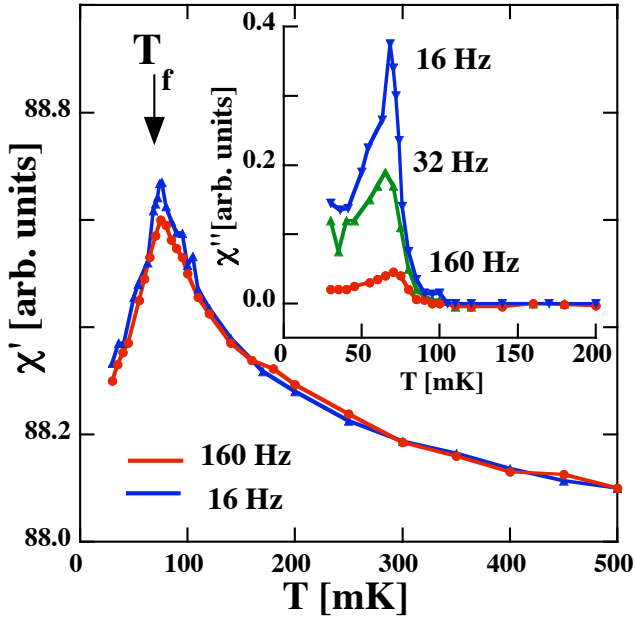}
\label{chimK}
\end{figure}
First of all, our ac-susceptibility measurements confirm the sharp anomaly in $\chi(T)$ in the form of a "cusp".  The temperature dependencies of both, the in-phase $\chi'$ and out-of-phase $\chi''$ components display peaks near 76 mK (Fig. \ref{chimK}), which is slightly lower than the reported value\cite{Gavilano05}. Although $\chi'$ has only a very weak frequency dependence, the magnitude and shape of $\chi''$ (Fig.~\ref{chimK} inset) are strongly frequency dependent. In particular, the temperature where the maximum of  $\chi'$ occurs decreases slightly with decreasing frequency. These observations hint for a slow dynamics of the V$^{4+}$ moments reminiscent of spin-glass-like processes with a freezing temperature $T_f$=76 mK.\\
Although the anomaly in $\chi(T)$ at $T_f$ signals a change in the magnetic properties of Na$_2$V$_3$O$_7$, the state below $T_f$ is not long-range magnetically ordered. This is consistent with 
powder neutron diffraction. The patterns collected at 40 mK and 1 K were identical, no magnetic reflections occurred at low temperature.\\  
Fig.~\ref{CpT} inset displays the low-temperature part of the magnetic specific heat $C_m$ obtained by subtracting the lattice contribution from the measured signal\cite{note1}. $C_m(T)$ shows no clear features near $T_{f}$. This important result shows that the number of degrees of freedom involved in the spin-freezing phenomenon must be rather small compared to the number of V$^{4+}$ ions. The second important feature is a broad Schottky-like feature near 0.4 K, which shifts to higher temperatures with applied field. Approximately 1/13 of the magnetic entropy is associated with this feature. It is worth to recall that there is also a broad feature at 5 K,  where at most 1/9 of the magnetic entropy is released as found previously~\cite{Gavilano05}.\\
\begin{figure}[tbh]
\caption{
Magnetic part of the specific heat divided by temperature, $C_m/T$, versus $T$ for different fields. Inset: $C_m$ versus $T$ below 1 K.}
\includegraphics[width=86mm,keepaspectratio=true]{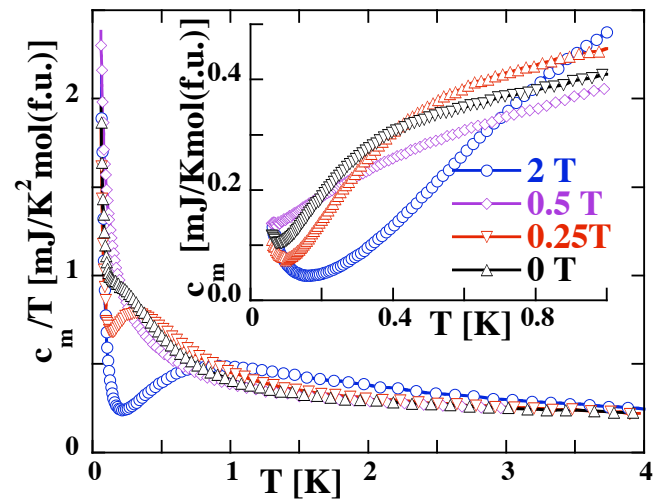}
\label{CpT}
\end{figure}
And finally, there is a  strong upturn in $C_{m}/T$ with decreasing temperature below 0.15 K (Fig.~\ref{CpT}). This upturn is well represented by $ C_N/T = D/T^3$. $D$ is of the order of 2$ \times 10^{-4}$  to 4$ \times 10^{-4}$ (J (mole f.u. K)$^{-1}$) and  $C_N$  is only weakly field dependent. 
Paramagnetic impurities (i.e. only weakly- or non- interacting magnetic ions) cannot be the origin of this feature because their contribution would be strongly suppressed by a field of 2 T, contrary to the experimental observations. The impurities, however, could come from a nuclear contribution to the specific heat due to i) nuclear quadrupolar energy levels and ii) nuclear Zeeman levels. \\First let's roughly estimate the contribution from nuclear quadrupolar energy levels.  Nuclei with spin $I$ and quadrupolar moment $eQ$ exposed to static electric-field gradients, where the largest component is denoted as $eq$, contribute to the specific heat (per mol of the considered nuclei)\cite{Phillips71}:
$C_N = D_Q \times T^{-2}$, with 
\begin{equation}
\frac{D_Q}{R} = \frac{1}{80} \frac{(2I+2)(2I+3)}{2I(2I-1)} \left( \frac{e^2qQ}{k_B} \right)^2
\label{C_Q}
\end{equation}
$R$ is the universal gas constant, $k_B$ is the Boltzmann constant and $e$ is the electron charge. The most promising candidates to provide a substantial contribution to the specific heat are the $^{51}$V nuclei which have\cite{Bruker07} $I = 7/2$, a large $Q = 5.2 \times 10^{-15}$ m$^2$  and low point symmetry.  From Eq.~\ref{C_Q} with  $D_{Q} = 4 \times 10^{-4}$ (J mole f.u. K$^{-1}$) it follows $eq \approx 4 \times 10^{23}$ V/m. This value is about two orders of magnitude too large compared with experimental NMR studies of other V compounds\cite{Ohama99}. Therefore, the nuclear quadrupolar levels seem to be improbable reason of our observed results.\\ 
Second we consider nuclear Zeeman levels.  Assume a fraction $f$ of magnetic V moments, which are at/near freezing. The direct contribution from these moments was already discarded previously. 
However, an indirect contribution via the hyperfine field coupling to the nuclei should also be taken into account. The nuclei are influenced by the on-site moment via the core polarization. This would result in a local field of the order of 100 T (per $\mu_B$ of onsite V moment)\cite{Ohama99}. The magnetic nuclear contribution to C$_N$ is 
$C_N = D_Z /T^{2}$, with 
\begin{equation}
\frac{D_Z}{R}= 3f  \frac{(2I+2)}{6I}\left( \frac{\gamma_N \hbar I H_l}{k_B} \right)^2.
\label{C_Z}
\end{equation}
This takes into account that there are 3 V ions per formula unit and assumes that only a fraction $f$ of V ions are magnetic.  $\gamma_N = 7.045\times 10^7$ (rad (sT)$^{-1}$T)\cite{Bruker07} is the nuclear gyromagnetic ratio of $^{51}$V and $H_l$ is the local magnetic field. With $I = 7/2$ and $H_l = 100$ T this leads to $f = 0.001$. We conclude that a tiny amount of "frozen" V moments acting on their nuclei via the hyperfine field would explain the observed C$_N(T)$ at low temperatures.\\

\subsubsection{Decay of remanent magnetization} \label{24}

\begin{figure}[tbh]
\caption {Temperature and frequency dependence of  the normalized remanent magnetization $m_{rem} =\frac{M_{t} - M_{\infty}}{M_{0} - M_{\infty}}$. Top: the same cooling field of 100 Oe has been used and measurement performed at different temperatures. Inset:  $M_{rem}$ [arb. units] versus $T$. Bottom: different cooling fields of 100 and 10 Oe have been used and measurement performed at 40 mK.
Inset: A versus $T^\beta$.}
\includegraphics[width=0.45\textwidth]{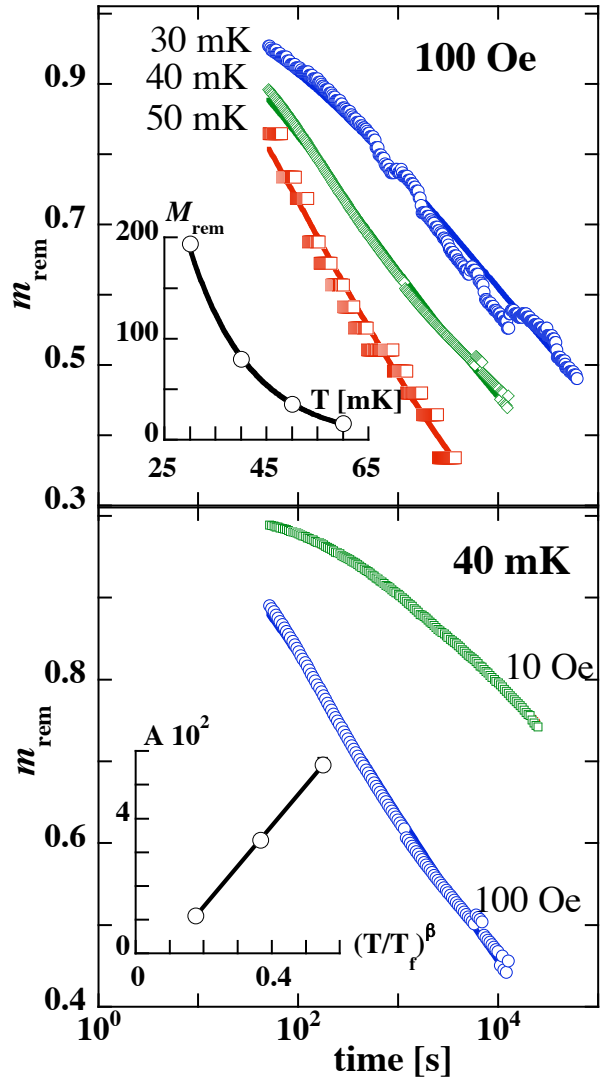}
\label{Mrem}
\end{figure}

The dynamics of the V moments on a long-time scale at temperatures below $T_f$ was probed by measuring the decay of the remanent magnetization $M_{rem}= M_{0}-M_{\infty}$. We recall that in these experiments the sample is cooled to a given temperature T$<$T$_{f}$ in a fixed external magnetic field H =100 Oe. After a waiting time of about 10h, H  is turned to zero at t=0 and the time evolution of the magnetization $M_{t}$ is monitored, keeping T fixed. The decay of $M_{t}-M_{\infty}$ does not follow a simple functional form and is extremely slow (Fig.~\ref{Mrem}) indicative of a spin-glass like behaviour. Characteristic time scales are of the order of 10$^4$ sec, which we define as the time where $M_{t}-M_{\infty}$ reaches a fraction  $1/e$ of its initial value. As shown in the inset of Fig.~\ref{Mrem} top, $M_{rem}$ grows exponentially with decreasing temperature.\\
For different spin-glass systems the magnetization decays following different functional forms. 
The functional form of $M_{t}-M_{\infty}$ is determined by the distribution of barriers that the system should overcome to reach the equilibrium. Our data are well characterized by the so-called "enhanced power law" given by~\cite{Hemmen85,Hemmen86}:
\begin{equation}
m_{rem} = \frac{M_{t} - M_{\infty}}{M_{0} - M_{\infty}} = \exp\{-A[\ln( \omega t ) ]^{\beta} \}
 \label{Model_M_t}
\end{equation}
This functional form is rather general, and it reduces, for instance,  to a "logarithmic" type-decay for the particular case of $\beta \approx 1$, $A<<1$ and $t$ not too large.\\
The underlying model assumes that as temperature decreases there is a spontaneous formation of clusters of correlated spins with non-zero total spin, which grow randomly and independently from each other leaving a "fluid" of  individual spins in between. To reach the thermal equilibrium a spin-reorientation of these clusters  must occur, and barriers, which depend on the cluster volume V, must be overcome to reach thermal equilibrium. This yields a hierarchy of relaxation times and results in the enhanced power-law for the decay of the magnetization.\\
The above form of the decay does not depend on the mechanism that drives the relaxation, which may be either quantum tunneling (QT) or thermal activation (TA). In addition the three parameters describing relaxation ${ A, \omega, \beta}$ do not independently vary with temperature. For instance,  the attempt frequency $\omega$ may be assumed  $T-$independent. In case of quantum tunneling the parameters  $A$ and $\beta$ are also $T-$independent, resulting in a temperature independent relaxation. This is clearly not the case in our experiments (Fig. ~\ref{Mrem}). In the case of thermal activation $\omega$ is $T-$independent, but $A$ and $\beta$ are correlated; namely, $A \propto (T/T_f)^\beta$ ~\cite{Hemmen86}. To a good approximation $A$ versus $T^\beta$ deduced from our fits (inset of Fig.~\ref{Mrem} bottom) is a straight line. Therefore, our data suggest a classical type of spin-glass phenomena driven by thermal activation, even at the lowest temperatures.\\
Our observations are not consistent with the 'super paramagnetic' picture describing the Mn$_{12}$\cite{Luis97}, V$_{15}$\cite{Chiorescu03}, Fe$_{13}$\cite{Slageren06} cluster systems. There, a well defined and unique magnetic barrier  leads to a single relaxation time resulting in a single exponential decay.\\ 

\section{{\it ab initio} DFT calculations} \label{3}

We have performed  {\it ab initio} DFT calculations for
the new crystal structure determined in this work (see Table II)
and applied the NMTO-based downfolding method in order
to obtain the effective vanadium-vanadium hopping matrix
elements. We considered a crystal structure with 100$\%$
occupation of the Na positions. The analysis of the Na 
density of states (DOS) contribution
near the Fermi level shows 
that the effect of Na is almost negligible concerning the
V hybridizations, and therefore significant changes
of the electronic structure are not to be expected by reducing
the Na content  in the calculations.
Variations in the hopping integral
 determination between 
the crystal structure proposed by Millet {\it et al.}\cite{Millet99,Dasgupta05} and the present
one are only observed in (i) the 
1$^{st}$ and 2$^{nd}$ intra-ring hopping parameters which are now
almost identical, enforcing the frustration effects in this
system, and (ii) the inter-ring hopping 
parameters are slightly larger
than those obtained previously, but remain still weaker
than the intra-ring hopping parameters. The analysis of the orbital
orientation in the 2$^{nd}$ intra-ring and 2$^{nd}$ inter-ring
paths shows that, eventhough distances and angles are
similar for both cases, the inter-ring hybridizations are
of $dd-\delta$ type explaining the weakness of the
hopping integral in contrast to the mixed $dd-\sigma$ $dd-\pi$ nature of the
intra-ring orbital hybridization. 
Still, the exchange coupling constants between $V_i$ and
$V_j$,  $J_{ij}$
are very sensitive to
ferromagnetic $J_{ij}^{FM}$ and antiferromagnetic  
 $J_{ij}^{AFM}$ contributions, $J_{ij}=   J_{ij}^{FM} + J_{ij}^{AFM}$.
 Calculation of $J_{ij}^{AFM}$
due to the hybridization of V$d_{xy}$ orbitals,
shows that
the antiferromagnetic intra-ring exchange interactions are a factor of four
larger than the antiferromagnetic inter-ring exchange interactions.
The ferromagnetic contributions  $J_{ij}^{FM}$,  which can be estimated following
the Kugel-Khomskii model\cite{Kugel82,Mazurenko06} are, on the contrary, of
the same order of magnitude for intra- and inter-ring couplings,  
   so that in conclusion,
the intra-ring interactions $J_{intra}$ are dominantly antiferromagnetic,
while the inter-ring interactions $J_{inter}$ have important ferromagnetic
components
 as pointed out also  in Ref.~\onlinecite{Mazurenko06}.

\section{Summary}\label{4}
We presented new experimental results which allow to advance the understanding of the puzzling properties of the nanotube system ${\tubex}$. 
Revisited crystal structure belongs to the $P\overline{3}$ space group, while the apparent $\overline{3}m$ symmetry is due to twinning. The bucked ring of nine V atoms, is the same as in the model of Millet {\it et al.}\cite{Millet99}, but the orientation of rings in the nanotube is different. The A'-B two-ring stacking unit is the basic structural building block.\\
We found off-stoichiometry in the sodium composition that might introduce randomness into the exchange couplings or presence of up to 5.7\% of V$^{5+}$ defects.\\
The lattice compression with decreasing temperature is very anisotropic, with several anomalous points that correlate with previous NMR and optical measurements. With decreasing temperature the dominant tendency is a reduction of distances directed mostly along the tube and an increase of those having substantial xy-component. No experimental support of a structural transition, spin dimerization and Na rattling was found. \\ 
In the neutron scattering experiments we observed a weak magnetic signal in the energy window of our experiment from -20 to 9 meV, implying that the majority of magnetic excitations are located above 20 meV. This is consistent with specific heat data, which indicate that at most 1/9 of the total magnetic entropy is released below 20 K, and with the susceptibility data, which monitor one order of magnitude moment reduction below 100 K. The magnetic scattering does not have a pronounced Q-dependence expected for spin clusters and, therefore, does not support the models of dimer formation or presence of isolated nine-member rings. The data might be explained, however, by the existence of 
clusters with different size and couplings or/and
significant inter-cluster couplings smearing out oscillations in $S(Q, \int \omega)$. We note, however, that these scenarios leave unexplained the prominent peak in the temperature dependence of the spin-lattice relaxation rate around 100 K\cite{Gavilano05}. That peak was used as part of the evidence that spin dimerization may occur in that temperature range. But our present findings show that it is improbable.\\

\begin{figure}[tbh]
\caption {Temperature dependence of inverse susceptibility obtained from our experiment (symbols) and calculated  (solid line) for an isolated nine-member ring with $J_1$= -160 K and $J_2$= -90 K. Inset: energy distribution of 512 levels of
the ring as a function of the total spin S. Calculations were performed using the alps program\cite{alps}.}
\includegraphics[width=0.45\textwidth]{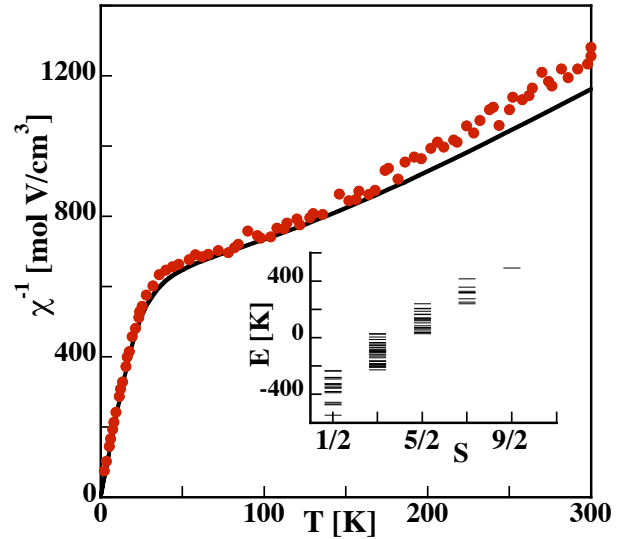}
\label{chiso}
\end{figure}

Our present and previous experimental observations allow us to suggest the following picture: The ${\tubex}$ system consists of segmented S=1/2 spin nanotubes with strong AF intra-ring but also significant inter-ring couplings. We speculate that segmentation happens due to the defects occurring because of the deviations from ideal stoichiometry. \\
Though the dominant feature of susceptibility, the upturn at 100 K, can be well reproduced by an isolated nine-member ring with $J_1$= -160 K and $J_2$= -90 K (Fig.~\ref{chiso}), other experimental observations on ${\tubex}$, i.e. neutron magnetic scattering, object a picture of isolated nine-member rings and require the presence of 
 inter-ring couplings or more complex spin objects. 
The susceptibility is dominated by strong intra-cluster couplings and is not a sensitive measure of the weaker inter-cluster couplings. A similar situation happens in coupled systems of triangles\cite{Schnack04} and tetrahedra\cite{Johnsson00, Zaharko04}.\\
{\it ab initio} calculations point to larger intra-ring
than inter-ring hopping integrals. The evaluation of the exchange
integrals  suggest strong antiferromagnetic inter-ring interactions and
moderate intra-ring interactions mostly of  ferromagnetic nature.\\
As there is a small magnetic moment at low T, we infer that ${\tubex}$ belongs to the class of odd-legged S=1/2 spin tubes with a gapless excitation spectrum and main spectral weight located at high T. 
As shown by L\"{u}scher et al\cite{Luscher04} frustrated inter-ring couplings might lead to the absence of a spin gap, which otherwise should be present in odd-legged tubes.
A single crystal inelastic neutron scattering experiment is needed to verify our hypothesis, to obtain the inter-ring couplings and to explain the low-temperature features at 0.4 and 5 K in specific heat.\\
The basic question, whether the ground state is magnetic or non-magnetic, still cannot be answered unambiguously. 
The origin of the 76 mK anomaly and of the observed slow dynamics in the mK range is not immediately evident. Our data strongly suggest 
a classical "spin-freezing" type phenomenon.  But, given the small amount of spins involved (of the order of 0.1\%) it is difficult to ascertain whether this behaviour is intrinsic to the ideal nanotube or it is induced by segmentation of nanotubes.\\
\begin{table}
\caption{Details of single crystal x-ray diffraction data collection and structure refinement for ${\tubex}$ at three different temperatures.}
\label{tab1}
\begin{ruledtabular}
\begin{tabular}{llll}
T (K)&200&80&16\\
\\
a (\AA)&10.8827(5)&10.8747(3)&10.8748(1)\\
c (\AA)&9.5506(2)&9.5382(1)&9.5275(2)\\
V (\AA$^3$)&979.57&976.86&975.77\\
2$\theta_{max}(deg)$&59.75&59.81&42.99\\
$R_{int} \%$&0.0182&0.0153&0.0423\\
Refl. collected&10485&1744&689\\
Refl. unique&1744&1744&689\\
restraints&0&0&18\\
refined parameters&113&113&69\\
R, wR ($I\ge2\sigma(I)$) \%&1.24, 3.99&1.23, 3.78&1.91, 5.16\\
GooF on $F^2$&1.23&1.15&1.17\\
largest diff. peak max/min&0.33/-0.32&0.38/-0.33&0.31/-0.27\\
and hole/e$\cdot$\AA$^{-3}$&&&\\
\end{tabular}
\end{ruledtabular}
\end{table}

\begin{table}
\caption{Atomic coordinates $xyz$ and isotropic displacement parameters U$_{iso}(\cdot$10$^2$\AA$^2$) in $\tubex$ at 200 K. The space group is $P\overline{3}$, Z=6,
twin proportion is 49.90(9) and 50.10(9)\%. The Na1 and Na2 sites are occupied only 45.3(11) and 51.8(11) \%.
Atomic parameters at 80 K and 16 K can be obtained as supplementary material.}
\label{tab2}
\begin{ruledtabular}
\begin{tabular}{cccccc}
atom&site&x&y&z&U$_{iso}$\\
V1  &6$g$&0.31549(3)   &0.10972(3)   &0.40031(3) &0.354(7)\\
V2  &6$g$&0.31643(3)   &0.149623(3)  & 0.75026(4) &0.316(6)\\
V3  &6$g$&0.31391(3)   &0.18514(3)   &0.10094(3) &0.390(7)\\
Na1&1$a$&0   &0   &0     &0.8(1)\\
Na2&1$b$&0   &0   &0.5     &1.2(1)\\ 
Na3&2$d$&${\onethree}$   &${\twothree}$   &0.0701(1)  &0.79(2)\\
Na4&2$d$&${\onethree}$   &${\twothree}$    &0.4332(1)  &0.95(2)\\
Na5&6$g$&0.5074(1)   &0.4805(1)   &0.2492(1) &2.07(2)\\
O1  &6$g$&0.4780(1)   &0.1491(2)   &0.4380(1) &1.02(2)\\
O2  &6$g$&0.4889(1)   &0.2194(1)   &0.7517(2) &1.02(3)\\
O3  &6$g$&0.4760(2)   &0.3083(2)   &0.0636(1) &1.15(3)\\
O4  &6$g$&0.2475(1)   &0.9508(1)   &0.7091(1) &0.59(2)\\
O5  &6$g$&0.2432(2)   &0.0664(1)   &0.9329(1) &0.65(2)\\
O6  &6$g$&0.2526(1)   &0.1701(2)   &0.5668(1) &0.67(2)\\
O7  &6$g$&0.2715(1)   &0.2987(1)   &0.7911(1) &0.52(2)\\
\end{tabular}
\end{ruledtabular}
\end{table}

\begin{table}
\caption{Anisotropic displacement parameters  U$_{ij}(\cdot$10$^2$\AA$^2$) in $\tubex$ at 200 K. }
\label{tab2b}
\begin{ruledtabular}
\begin{tabular}{ccccccc}
atom&U$_{11}$&U$_{22}$&U$_{33}$&U$_{23}$&U$_{13}$&U$_{12}$\\
V1  &0.36(1) &0.40(1) &0.30(1) &-0.044(9) &-0.07(1) &0.19(1)\\
V2  &0.28(1) &0.39(1) &0.25(1) &-0.023(9) &-0.02(1) &0.15(1)\\
V3  &0.42(1) &0.46(1) &0.31(1) &0.011(9) &0.06(1) &0.23(1)\\
Na1&0.7(2) &0.7(2) &1.2(2) &0 &0 &0.34(8)\\
Na2&1.1(1) &1.1(1) &1.7(2) &0 &0 &0.54(7)\\
Na3&0.79(3) &0.79(3) &0.80(6) &0 &0 &0.39(2)\\
Na4&1.00(3) &1.00(3) &0.87(6) &0 &0 &0.50(2)\\
Na5&1.16(4) &1.56(4) &2.51(4) &-0.40(3) &-0.52(3) &-0.05(3)\\
O1  &0.69(6) &1.39(7) &1.08(6) &-0.42(5) &-0.25(5) &0.58(5)\\
O2  &0.54(6) &1.30(7) &0.89(5) &-0.12(5) &-0.03(5) &0.22(5)\\
O3  &0.69(6) &1.31(7) &1.11(6) &-0.16(5) &0.16(5) &0.25(5)\\
O4  &0.80(5) &0.41(5) &0.34(5) &-0.04(4) &0.03(4) &0.14(5)\\
O5  &0.99(6) &0.48(6) &0.38(6) &-0.04(4) &0.11(4) &0.31(5)\\
O6  &1.23(6) &0.77(6) &0.33(6) &-0.04(4) &0.04(5) &0.74(6)\\
O7  &0.65(5) &0.67(6) &0.38(5) &-0.02(4) &0.02(4) &0.43(5)\\
\end{tabular}
\end{ruledtabular}
\end{table}

\begin{table}
\caption{Shortest V-V distances (\AA) at 200 K, 80 K, 16 K and difference $\Delta$d between the values at 16 K and 200 K. The notations of Figure~\ref{fig2}d are used.}
\label{tab3}
\begin{ruledtabular}
\begin{tabular}{lcccc}
V-V distance&200 K&80 K& 16 K&$\Delta$d\\
&\multicolumn{2}{c}{1$^{st}$ intra-ring}&&\\
11-31&2.9770(4)&2.9648(4)&2.955(1)&-0.022(1)\\
21-32&2.9851(4)&2.9908(4)&2.996(1)&+0.011(1)\\
11-21&2.9917(5)&2.9976(4)&3.001(1)&+0.010(1)\\
&\multicolumn{2}{c}{2$^{nd}$ intra-ring}&&\\
21-31&3.6109(4)&3.6008(4)&3.596(1)&-0.015(1)\\
11-22&3.6588(5)&3.6635(4)&3.663(1)&+0.004(1)\\
&\multicolumn{2}{c}{1$^{st}$ inter-ring}&&\\
11-23&3.3696(5)&3.3670(4)&3.365(1)&-0.004(1)\\
24-31&3.3732(5)&3.3716(4)&3.369(1)&-0.005(1)\\
&\multicolumn{2}{c}{2$^{nd}$ inter-ring}&&\\
31-33&3.5447(4)&3.5436(4)&3.545(1)&0.000(1)\\
11-12&3.5694(5)&3.5760(4)&3.578(1)&+0.009(1)\\
\end{tabular}
\end{ruledtabular}
\end{table}

\begin{table}
\caption{V-O-V angles (deg) at 200 K, 80 K, 16 K associated with the shortest V-V distances. The notations of Figure~\ref{fig2}d are used.}
\label{tab4}
\begin{ruledtabular}
\begin{tabular}{lccc}
V-O-V angle&200 K&80 K& 16 K\\
&\multicolumn{2}{c}{1$^{st}$ intra-ring}&\\
11-O7-31&98.01(6) &97.28(5) &97.1(1)\\
11-O4-31&99.56(6) &99.26(5) &98.9(1)\\
21-O4-32&99.13(6)&99.55(5)&99.9(1)\\
21-O5-32&101.31(6)&101.39(6)&101.4(2)\\
11-O7-21&98.19(5)&98.24(5)&98.4(1)\\
11-O6-21&101.86(6)&102.07(6)&102.3(2)\\
&\multicolumn{2}{c}{2$^{nd}$ intra-ring}&\\
21-O7-31&136.15(7)&135.41(7)&135.2(2))\\
11-O4-22&142.50(7)&143.41(7)&143.9(2)\\
&\multicolumn{2}{c}{1$^{st}$ inter-ring}&\\
11-O6-23&119.16(7)&119.02(7)&119.0(2)\\
24-O5-31&119.56(7)&119.48(6)&119.3(2)\\
&\multicolumn{2}{c}{2$^{nd}$ inter-ring}&\\
31-O5-33&132.20(7)&132.00(7)&132.0(2)\\
11-O6-12&133.49(7)&133.61(7)&133.5(2)\\
\end{tabular}
\end{ruledtabular}
\end{table}

\section{Acknowledgments}
The work was partially performed at SINQ, Paul Scherrer Insitute, Villigen, Switzerland, at the ILL reactor, and at the Swiss-Norwegian Beamlines, ESRF, Grenoble, France. The work at MPICPfS was carried out within the DFG project MI1171/1-1.
 The work at ITP Frankfurt was carried out
within the DFG project SFB/TR49.
 We thank for the expert assistance by Dr.  F. Gozzo, Swiss Light Source and Dr. L. Keller, SINQ, PSI.


\begin{thebibliography}{99}

\bibitem[*]{now} e-mail: Oksana.Zaharko@psi.ch
\bibitem{Millet99} P. Millet, J. Y. Henry, F. Mila, J. Galy J. Solid State Chem. {\bf 147}, 676 (1999).
\bibitem{Gavilano03} J. L. Gavilano, D. Rau, S. Mushkolaj, H. R. Ott, P. Millet, F. Mila, Phys. Rev. Lett. {\bf 90}, 167202 (2003).
\bibitem{Gavilano05} J. L. Gavilano, E. Felder, D. Rau, H. R. Ott, P. Millet, F. Mila, T. Cichorek, A. C. Mota, Phys. Rev. B  {\bf 72}, 64431 (2005).
\bibitem{Whangbo00} M.-H. Whangbo and H.-J. Koo Solid State Comm. {\bf115}, 675 (2000).
\bibitem{Dasgupta05} T. Saha-Dasgupta, R. Valent\'{i}, F. Capraro, C. Gros, Phys. Rev. Lett. {\bf95}, 107201 (2005).
\bibitem{Mazurenko06}  V.V. Mazurenko, F. Mila, V. I. Anisimov, Phys. Rev. B {\bf73}, 14418 (2006).
\bibitem{Niitaka02} S. Niitaka, K. Yoshimura, A. Ikawa, K. Kosuge, J. Phys. Soc. Jpn {\bf71}, 208 (2002) Suppl.
\bibitem{note0} The sample was stored in He atmosphere as it was unstable in air. Unfortunately a small amount of impurity NaVO$_3$ could not be avoided. However it contains nonmagnetic V$^{5+}$ and does not affect our findings on the magnetic properties of ${\tube}$.\\
\bibitem{SHELX} G. M. Sheldrick SHELXS97 and SHELXL97. Programs for the solution and refinement of crystal
structures; University of G\"{o}ttingen: G\"{o}ttingen, Germany, 1997.
\bibitem{fullprof} J. Rodriquez-Carvajal, Physica {\bf192B}, 55(1993).
\bibitem{Stewart2008} J. R. Stewart, P. P. Deen, K. H. Andersen, H. Schober, J. F. Barthelemy, 
J. H. Hillier, A. P. Murani, T. Hayes, B. Lindenau and P. Hoghoj, J. Appl. Cryst. to be published.
\bibitem{Vertesi06} T. V\'{e}rtesi, E. Bene, Phys. Rev. B {\bf73}, 134404 (2006).
\bibitem{Scharpf} O. Sch\"{a}rpf, H. Capellmann, Phys. Stat. Sol. (a) {\bf 135}, 359 (1993).
\bibitem{Furrer79} A. Furrer, H. U. G\"{u}del, J. Magn, Magn. Mater {\bf14}, 256 (1979).
\bibitem{Woihe} H. Weihe, Department of Chemistry, University of Copenhagen, Denmark.
\bibitem{note1} The lattice contribution $C_{lat}$ was estimated using the Debye model with $\Theta_{D} $  = 125 K (the contribution from the optical modes is negligible in our experiments).  
\bibitem{Choi02} J. Choi, J. L. Musfeldt, Y. J. Wang, H.-J. Koo, M.-H. Whangbo J. Galy, P. Millet, Chem. Mater. {\bf 14}, 924 (2002).
\bibitem{Phillips71} N. E. Phillips, CRC Critical Reviews in Solid State Sciences, p 467 (1971).
\bibitem{Bruker07} In "{\it NMR properties of Selected Isotopes"}, p. 10, Bruker Almanac, 2007.
\bibitem{Ohama99} T. Ohama, H. Yasuoka, M Isobe and Y Ueda,  Phys. Rev. B {\bf 59}, 3299 (1999).
\bibitem{Fischer91} Spin glasses, edited by K. H. Fischer and J. A. Hertz (Cambridge University Press, Cambridge, England, 1991); 
J. A. Mydosh Spin glasses: An Experimental Introduction (Taylor and Francis, London, 1993). 
\bibitem{Hemmen85} J. L. van Hemmen, A S\"{u}to, Z. Phys. B - Condensed Matter {\bf 61}, 263 (1985).
\bibitem{Hemmen86} J. L. van Hemmen, G. J. Nieuwenhuys, Europhys. Lett. {\bf 2}, 797 (1986).
\bibitem{Luis97} F. Luis, J. Bartolom\'{e}, J. F. Fern\'{a}ndez, J. Tejada, J. M. Hern\'{a}ndez, X. X. Zhang, R. Ziolo, Phys. Rev. B {\bf 55}, 11448(1997).
\bibitem{Chiorescu03} I. Chiorescu, W. Wernsdorfer, A. M\"{uller}, S. Miyashita, B. Barbara,  Phys. Rev. B {\bf 67}, 020402(R)(2003).
\bibitem{Slageren06} J. van Slageren, P. Rosa, A. Caneschi, R. Sessoli, H. Casellas, Y. V. Rakitin, L. Cianchi, F. Del Giallo, G. Spina, A. Bino, A.-L. Barra, T. Guidi, S. Carretta, and R. Caciuffo, Phys. Rev. B {\bf 73}, 014422(2006).
\bibitem{Kugel82} K. I. Kugel and D. I. Khomskii, Sov. Phys. Usp. {\bf 25}, 231 (1982).
\bibitem{magpack} J. J. Borr\'{a}s-Almenar, J. M. Clemente-Juan, E. Coronado, B.S. Tsukerblat Inorg. Chem. 38(1999)6081. 
\bibitem{Schnack04} J. Schnack, H. Nojiri, P. K\"{o}gerler, G. J. T. Cooper, L. Cronin, Phys. Rev. B {\bf 70}, 174420(2004).
\bibitem{Johnsson00} M. Johnsson K. W. T\"{o}rnroos, F. Mila, P. Millet, Chem. Mater. {\bf12}, 2853 (2000).
\bibitem{Zaharko04} O. Zaharko, A. Daoud-Aladine, S. Streule, J. Mesot, P. J. Brown, and H. Berger, Phys. Rev. Lett. {\bf93}, 217206 (2004).
\bibitem{Luscher04} A. L\"{u}scher, R. M. Noack, G. Misguich, V. N. Kotov, F. Mila,  Phys. Rev. B {\bf 70}, 060405(R)(2004).
\bibitem{alps} A. F. Albuquerque, F. Alet, P. Corboz, P. Dayal, A. Feiguin, S. Fuchs, L. Gamper, E. Gull, S. Guertler, A. Honecker, R. Igarashi, M. Koerner, A. Kozhevnikov, A. Laeuchli, S.R. Manmana, M. Matsumoto, I.P. McCulloch, F. Michel, R.M. Noack, G. Pawlowski, L. Pollet, T. Pruschke, U. Schollwock, S. Todo, S. Trebst, M. Troyer, P. Werner, S. Wessel, J. of Magn. and Magn. Materials 310, 1187 (2007).

\end{thebibliography}
\end{document}